# A comparison study between the properties of $BHBT_2$:$PC_{71}BM$ composite nanotubes and its bulk-heterojunction

**Muhamad Doris[1,2*], Azzuliani Supangat[2], Teh Chin Hoong[2], Khaulah Sulaiman[2] and Rusli Daik[3]**

[1] Department of Electrical Engineering, Universitas Trisakti, Indonesia
[2] Low Dimensional Materials Research Centre, Department of Physics, University of Malaya, Malaysia
[3] School of Chemical Sciences and Food Technology, Faculty of Science and Technology, Universiti Kebangsaan Malaysia, Malaysia

## Abstract

In this study, we investigate the morphological, structural, and optical properties of both bulk-heterojunction and composite nanotubes that are composed of thiophene-based small molecules 1,4-bis (2,2'-bithiopen-5-yl)2,5-dihexyloxybenzene ($BHBT_2$) and fullerene [6,6]-phenyl $C_{71}$ butyric acid methyl ester ($PC_{71}BM$). Mix-blended and template-assisted methods were used to fabricate the bulk-heterojunction and the composite nanotubes, respectively. A single material of $BHBT_2$ thin films and nanotubes fabricated via spin-coating and template-assisted methods, respectively, were also studied. Two different formations between the bulk-heterojunction and the composite nanotubes were compared to elaborate on their advancement in properties. The absorption spectra of the $BHBT_2$:$PC_{71}BM$ bulk-heterojunction and the composite nanotubes were found to have fallen within the ultraviolet-visible (UV-vis) range. Field emission scanning electron microscope (FESEM) and high-resolution transmission electron microscope (HRTEM) images showed that the carbon-rich of $PC_{71}BM$, had successfully infiltrated into the $BHBT_2$ nanotube to form the $BHBT_2$:$PC_{71}BM$

* Corresponding author: m.doris@trisakti.ac.id
[1] Department of Electrical Engineering
Universitas Trisakti, Jakarta, 11440, Indonesia
[2] Low Dimensional Materials Research Centre, Department of Physics
University of Malaya, Kuala Lumpur, 50603, Malaysia
[2] School of Chemical Sciences and Food Technology, Faculty of Science and Technology, Universiti Kebangsaan Malaysia, 43600 UKM Bangi, Selangor, Malaysia

composite nanotubes. However, due to the inconsistent uniformed informal distribution of the infiltration, the charge carrier transfer is seen to be better in the bulk-heterojunction rather than in the composite nanotubes.

**Background**

Organic materials have been of interest to many researchers due to their extraordinary properties such as flexibility, abundancy of resources, and low production cost. Initially, the organic materials are only used as electric insulator of electronic devices. However, this has changed after the discovery of the metallic behavior of polyacetylene (PA) by Shirakawa et.al [1]. This discovery has lead to the emergence of other conducting organic materials such as polythiophene (Pth), polypyrrole (PPY), polyaniline (PANI), polycarbazole (PCz), polyparaphenylene vinylene (PPV), poly(3,4-ethylene dioxythiophene) (PEDOT), and polyfuran (PF) [2-9] that to date have been used in numerous electronic devices applications. In addition, many other new organic materials have recently been discovered and used in applications involving electronic nanodevices (field effect transistor, light emitting diode) [10, 11], sensors (chemical, gas, optical, biosensor) [12-15], energy storage systems (solar cells, fuel cells, supercapacitors) [16-18], microwave absorption and electromagnetic shielding [19], and biomedical procedures (drug delivery, protein purification, tissue engineering, neural interfaces, actuators) [20-24]. These materials and applications are fabricated based on polymer which have different properties over its small molecules.

The small molecules based materials are promising as they have the properties of viscoelasticity, toughness, crystallinity, and optical when compared to polymer based materials [25-30]. In this study, we examine a novel Thiophene-based small molecules in which the Thiophene is a heterocyclic compound that consists of a five-membered ring that has properties still seen as reliable and remarkable to date. Furthermore, Thiophene-based small molecules have extraordinary properties that may compete with its polymer in terms of enhancing the efficiency of perovskite-based solar cells and small molecules organic solar cells [28, 31]. Some applications on small molecules organic solar cells [28, 30-32], inorganic-organic hybrid solar cells [33], and field effect transistor [34-36] have been realized from the thiophene-based small molecules. Therefore, intensive studies of thiophene-based small molecules are as crucial as studies of its thiophene-conjugated polymer.

Thiophene-based small molecules, namely 1,4-bis (2,2'-bithiopen-5-yl)2,5-dihexyloxybenzene ($BHBT_2$) is a pentamer that consists of four thiophene rings and one benzene ring that binds the two hexyl molecules and thiophene. $BHBT_2$ has a profound solubility particularly in organic solvents such as chloroform, dichloromethane, and tetrahydrofuran. In addition, it is easily purified through column chromatography and recrystallization techniques. The terminal bithiophene groups in a $BHBT_2$ pentamer are expected to provide good stability and excellent charge transport properties in the pentamer backbone. Therefore, the coupling of dihexyloxy-p-phenylene moiety with terminal bithiophene groups could improve the pi-electron conjugation path in the pentamer backbone without affecting their optical and electrochemical properties [50].

In this work, one of the key explorations of $BHBT_2$ properties is on its nanostructures. Nanostructures of small molecules such as nanotubes, nanorods, nanowires, and nanoflowers have been commonly fabricated using the template assisted-method [37-41]. However, study of $BHBT_2$ small molecules nanostructures upon the template assisted-method have never been examined to date. The template used is based on alumina oxide material that consists of orderly nanopores to infiltrate a solution, which in turn molds the nanostructured materials into shape. Formation of nanostructures is important to be studied due to its exceptional properties over the planar structures. For instance, an active material in organic solar cells that is fabricated into nanostructures with the size that corresponds to visible wavelength can enhance the optical performance of the material [42]. In addition, nanostructuring an active material of solar cells into nanorods, nanotubes, and nanowires increases the surface area of the active layer allowing it to absorb more photons that results in higher efficiency [43, 44]. The template assisted-method is a low cost technique to produce nanostructured materials where the properties of the materials can easily be controlled [37, 45, 46]. The types of nanostructures are affected by the parameters of solution that infiltrate into the template's pores. Molecular weight, solution viscosity, solution concentration, infiltration mechanism, and drying process are some of the parameters that influence the formation of nanostructures [39, 40]. Template-assisted method also can be used to fabricate the donor-acceptor system of p-n junction composite nanotubes [47]. Unlike the well-known donor-acceptor system of bulk-heterojunction, the composite nanotubes that are composed of an ordered array of p-n core-shell junctions could accommodate well to the higher carriers mobility. The p-n core-shell junction is widely used in applications such as drug delivery and optoelectronic [48, 49]. As reported previously, a p-n core-shell junction nanowire that was applied to inorganic solar cells have been seen to have improved the efficiency performance of the device [44]. On top of that, the p-n core-shell junction of composite nanotubes could also enhance the charge carrier transfer when compared with its bulk-heterojunction [47].

In this study, the p-n core-shell junction of composite nanotubes has been successfully produced via the template assisted-method. Composite nanotubes of $BHBT_2$ and fullerene [6,6]-phenyl $C_{71}$ butyric acid methyl ester ($PC_{71}BM$) have also been fabricated and characterized. The characterizations are particularly emphasized on the optical, morphological, and structural properties of $BHBT_2$:$PC_{71}BM$. The two different formations of bulk-heterojunction and composite nanotubes of $BHBT_2$:$PC_{71}BM$ are then studied. To the best of our knowledge, there have been no studies that have investigated the properties of thiophene-based small molecules $BHBT_2$. Therefore, studies on the p-n $BHBT_2$:$PC_{71}BM$ properties may provide informative and useful knowledge.

## Methods

Thiophene-based small molecules, 1,4-bis (2,2'-bithiophene-5-yl)2,5-dihexyloxybenzene ($BHBT_2$) were synthesized and used directly as needed. As shown in Figure 1a, $BHBT_2$ contains a single benzene with four thiophene rings flank the benzene ring on each side. 5 mg of $BHBT_2$ was dissolved into the 1 ml of chloroform where the solubility limit of $BHBT_2$ within chloroform is 59 mg $ml^{-1}$ [50]. A similar concentration of 5 mg/ml was applied to [6,6]-phenyl $C_{71}$ butyric acid methyl ester ($PC_{71}BM$), by dissolving it in chloroform. The volume ratio of 1:1 was used to produce the $BHBT_2$:$PC_{71}BM$ bulk-heterojunction and composite nanotubes. A porous alumina template with 20 nm and 60 μm of pores in diameter and thickness, respectively, was purchased from Whatman Anodisc and was utilized to fabricate the nanostructures. Prior to the infiltration, the templates were cleaned by immersing them in acetone under sonication for 15 min, and then rinsed using deionized water. Three different spin coating rates of 1000, 2000, and 3000 rpm were used in 30 s. Prior to the spin coating process, the templates were first attached onto a glass slide using scotch tape on its right and left sides (Figure 1b). The scotch tape was used to ensure that the template would remain stuck on the glass slide during the spin coating process. Figure 1c shows the infiltrated $BHBT_2$ after and before the spin coating and dissolution process, respectively.

Figure 2 represents the schematic illustrations of the formation of the $BHBT_2$ nanotubes and the $BHBT_2$:$PC_{71}BM$ composite nanotubes. The porous alumina template was first cleaned under sonication of acetone (i). Prior to the spin coating process, 50 μL of $BHBT_2$ solution was dropped onto the cleaned template, which allowed the solution to infiltrate into the template (ii). To fabricate the $BHBT_2$:$PC_{71}BM$ composite nanotubes, 50 μL of $PC_{71}BM$ were dropped on top of the infiltrated $BHBT_2$ followed by the spin coating process (iii). Templates with infiltrated materials were then dried under room temperature.

These templates were stuck upside down on a copper tape (iv) before the dissolution of 6 h in 5 M of sodium hydroxide (NaOH) took place (v). In order to wash the remaining template thoroughly, it was rinsed several times using deionized water. Finally, the obtained nanotubes that remain stuck on the copper tape were ready to be characterized (vi). Schematic illustrations shown in Figure 2 were adapted from the sample preparation set up shown in Figure 3. It can be clearly seen that the infiltrated $BHBT_2$ presented with a yellow appearance (Figure 3a). The $BHBT_2$ nanotubes were able to retain their adhesion on the copper tape although the sample was washed for several times (Figure 3b). The $BHBT_2$:$PC_{71}BM$ composite nanotubes turned brownish due to the infiltration of the $PC_{71}BM$ (Figure 3c). As with the $BHBT_2$ nanotubes, the $BHBT_2$:$PC_{71}BM$ composite nanotubes were also be able to remain stuck upside down on the copper tape after several washing (Figure 3d).

Several equipment such as the spin coater model WS-650MZ-23NPP (Laurell Technologies Corp., North Wales, PA, USA), Field Emission Scanning Electron Microscope (FESEM) (Quanta FEG 450), High Resolution Transmission Electron Microscope (HRTEM) (Tecnai G2 FEI), Raman and photoluminescence (PL) spectroscopy (RENISHAW), UV-vis spectroscopy (Shimadzu UV-3101PC) and X-ray Diffraction Spectroscopy (XRD) were used in this study.

## Results and discussion

In this study, the properties of single material and composite materials comprised of $BHBT_2$ (p-type) and fullerene (n-type) are reported. $BHBT_2$ (1,4-bis (2,2'-bithiophene-5-yl)2,5-dihexyloxybenzene) is a novel small molecule with pentamer [50] and the advantage of high solubility with most solvents. By investigating its optical, morphological and structural properties, new applications on organic electronics devices can be realized to improve the device performance. Based on the X-ray Diffraction (XRD) measurement of $BHBT_2$ shown in Figure 4a, multiple peaks that represent the $BHBT_2$ pristine indicate that the $BHBT_2$ small molecule is a crystalline material [51]. Crystalline materials with periodic structure may provide a light management ability that can outperform the devices, for instance, the enhancement of photon absorption [52-54]. To elaborate on the properties of $BHBT_2$ small molecules further, incorporation of $BHBT_2$ and a p-type material of $PC_{71}BM$ is applied. Nanostructuring these materials into p-n composite may enrich knowledge in nano-morphology studies. In addition, the charge transfer between donor and acceptor materials can be better understood and digested when there is more information on the carrier's behaviors. Figure 4b shows the energy diagram of $BHBT_2$ and $PC_{71}BM$ with HOMO-LUMO of 1.96-4.65 eV and 3.94-5.93 eV, respectively. The HOMO-LUMO of donor-acceptor will enable the exciton (electron-hole) to dissociate at the interface [55].

### Optical properties studies

Figure 5a and Figure 5b show the absorption range of $BHBT_2$ thin films and the nanotubes spin-coated at the different rates of 1000, 2000, and 3000 rpm, respectively. Increasing the spin coating rate was applied to produce thinner films of the $BHBT_2$. A similar pattern of wavelength absorption is shown with the dissimilarity in absorption intensity where hypochromic effect have occurred in which decreasing the intensity of absorbance as effect of less materials that absorb photon. This can be understood that the differences in the thickness of material depends on the spin coating rate in which the thicker layer will absorb more photon. From the UV-vis absorption spectra, $BHBT_2$ thin films and nanotubes portrayed five significant absorption peaks. No single peak shifting either to left or right side was observed to have taken place, due to the different spin coating rates, apart from the changes in the absorption intensity of $BHBT_2$ nanotubes, which get higher at 350 nm when compared to its thin films. It was noticed that $BHBT_2$ favorably absorb photon only in the range of UV and visible light region. Due to its light absorption properties, $BHBT_2$ may potentially be applied as UV-vis photodetector or as a booster material of active layer in organic photovoltaic applications. Fabricating this material into highly ordered nanostructures may enhance its performance since the charge carrier's transfer mobility can be improved via the highly ordered structures [56, 57]. As shown in Figure 6a, $BHBT_2$ nanotubes reveal significantly higher absorption intensity in comparison to their thin films [56]. $BHBT_2$ nanotubes and thin films exhibit four shoulders with their significant peaks occuring at 335, 350, 414, 429, and 461 nm. The longer absorption wavelength of both $BHBT_2$ nanotubes and thin films are observed at 461 nm of Soret peak (B-band). This condition occurs due to movement of an electron dipole that corresponds to nonbonding-antibonding (n-$\pi$*) excitation among $BHBT_2$ molecules. A small part of UV light disclosed at around 320 - 350 nm is attributed to the bonding-antibonding ($\pi$ -$\pi$*) excitation. Generally, nanostructuring the $BHBT_2$ either in single material or composite has increased its absorption performance. Incorporating $BHBT_2$ with $PC_{71}BM$ slightly shifted the intense absorption and rendered the absorption range broader at 320-350 nm and 400-460 nm. As shown in Figure 6b, although the presence of $\pi$ -$\pi$* transition was observed in the $BHBT_2$ bulk-heterojunction, the better absorption was exhibited in the $BHBT_2$ composite nanotubes.

In an organic material, once a photon bombards onto a donor material, exciton, which is a pair of electron-hole that binds with each other, will be generated. In order to generate a current, this exciton will need to be dissociated by dislodging the electron from its binding energy. One condition that has to be fulfilled is the affinity electron requirement of each donor and acceptor materials. In our study, to investigate the potential application of $BHBT_2$ into a device, $BHBT_2$ and electron acceptor of $PC_{71}BM$ were mixed. It should be possible to create a charge transfer phenomena via the HOMO-LUMO configuration of both materials since the electron affinity of the acceptor material is bigger than that of the donor. Charge carriers transfer will only occur when the binding energy of exciton is lower than the electron affinity energy difference between the donor and acceptor materials [58].

The photoluminescence (PL) measurement was carried out to investigate the effectiveness of charge transfer between the donor and acceptor for both the bulk heterojunction and the composite nanotubes. The PL provides information on how well the exciton can reach the donor-acceptor interfaces [59]. Exciton (a pair of electron-hole) that has been generated by the donor material needs to be separated for the current extraction. Recombination of exciton will lead to the emission of radiative photon (radiative recombination) which is shown as intensity in the PL spectra [60]. Excitons that have successfully reached the donor-acceptor interfaces will be dissociated into hole and electron. This phenomenon is indicated by the lower intensity exhibited by the PL peak (quenching) compared to the curve of the radiative recombination which always exhibit the higher intensity. Quenching is attributed to the charge carriers transfer that had occurred between the donor and acceptor, which are carriers produced from the dissociated excitons at the donor-acceptor interfaces [59, 60]. As shown in Figure 7a, the quench phenomenon occurred due to the incorporation of $BHBT_2$ with $PC_{71}BM$, which allowed electrons from $BHBT_2$ to jump to the acceptor material. It is hardly to obtain the photoluminescence results for thin films that were spin-coated at 1000, 2000, and 3000 rpm at a concentration of 5 mg/ml based on the thickness matter. To solve this hindrance, drop-casting of the solution onto a glass substrate is applied in order to get the feasible films thickness. It is noticed that $BHBT_2$ thin films has a higher photoluminescence intensity compared to the $BHBT_2$ bulk-heterojunction and composite nanotubes. Mixing of two materials in the formation of bulk-heterojunction and composite nanotubes has gained almost totally quenching. Figure 7b shows the photoluminescence spectra of $BHBT_2$ bulk-heterojunction and composite nanotubes. $BHBT_2$ bulk-heterojunction got slightly quenched compared to its composite nanotubes. Charge transfer between $BHBT_2$ and $PC_{71}BM$ in the formation of bulk-heterojunction was more effective than the composite nanotubes. Based on the morphology view of bulk-heterojunction, the donor and acceptor will easily agglomerate and mix together in order to allow the electron to reach the donor-acceptor interface. Better quenching of $BHBT_2$:$PC_{71}BM$ bulk-heterojunction produced results which a contradict the results found by other studies [45, 47] where nanostructuring a material gets better quenching than bulk-heterojunctioning. The morphology of $BHBT_2$:$PC_{71}BM$ composite nanotubes could be responsible for the better quenching of bulk-heterojunction where it could be predicted that most of the $BHBT_2$:$PC_{71}BM$ composite nanotube interfaces were unevenly constructed.

**Structural properties studies**

Raman spectroscopy was applied to study the chemical and structural composition of materials. In its application, incident photons that interact with materials may lose or gain energy. The energy difference between the scattered photon and the incident photon is exactly similar to the energy difference in molecular vibration. Therefore, patterns of Raman spectra are generated from the molecular or lattice vibrations within the materials. A low frequency in the Raman shift corresponds to a low energy vibration of atoms which means that heavy atoms are held together with the weak bonds. On the other hand, a high frequency corresponded to light atoms which were held together with a strong bound [61]. Figure 8a and 8b show the Raman spectra of $BHBT_2$ thin films versus nanotubes and the Raman spectra of $BHBT_2$:$PC_{71}BM$ bulk-heterojunction versus composite nanotubes, respectively, with their Raman peaks tabulated in Table 1. Raman peak at 1445 $cm^{-1}$ shows a slightly different in intensity between $BHBT_2$ thin films and nanotubes indicating that the thiophenes ring was dominantly stretching in thin film compared to its nanotubes. The occurrence of thiophenes ring stretching supports the existence of $BHBT_2$molecular structure with four thiophene rings attached to one benzene ring. However, the incorporation of two materials resulted in shifting around 3 $cm^{-1}$ to the lower frequency. This shifting could be due to the effect of incidence energy that was divided and served to the two different molecules, which in turn resulted in a decrease in scattered energy [61]. Most of the Raman peaks of $BHBT_2$ nanotubes shifted to higher frequencies except for the peak at 1564 $cm^{-1}$ that shifted to lower a frequency for 2 $cm^{-1}$. However, no CH deformation occurred in $BHBT_2$ nanotubes. Ring breathing and ring vibration para-substituted benzene were formed from the incorporation of $BHBT_2$ and $PC_{71}BM$. These rings were found at 1188 $cm^{-1}$ and 1227 $cm^{-1}$, respectively. $BHBT_2$ composite nanotubes showed shifting to the higher frequencies of 1189 $cm^{-1}$, and 1231 $cm^{-1}$. Among the four Raman spectra ($BHBT_2$ thin films, $BHBT_2$ nanotubes, $BHBT_2$:$PC_{71}BM$ bulk heterojunction, and $BHBT_2$:$PC_{71}BM$ composite nanotubes), $BHBT_2$ nanotubes and $BHBT_2$:$PC_{71}BM$ composite nanotubes had somewhat a similar intensity although the highest peak intensity was dominated by $BHBT_2$ thin films.

**Formation of nanostructured composite**

$BHBT_2$ nanotubes were synthesized via the template-assisted method. Figure 9a-c show the FESEM images of $BHBT_2$ nanotubes obtained from the three different spin coating rates of 1000, 2000, and 3000 rpm. Generally, the nanostructures created bundles of nanotubes by collapsing their tips with each other instead of growing aligned as a single nanotube. Formation of nanotubes bundles could be due to the presence of attractive forces (van der Waals interactions) between the nanotubes [40]. At spin coating rate of 1000 rpm, the morphology of $BHBT_2$ nanotubes were produced inconsistently with some of the tubes longer than others. However, with the increase of the spin coating rate to 2000 rpm, the homogeneous growth of $BHBT_2$ nanotubes was attained. A further increase to 3000 rpm, resulted in a thicker base layer and denser nanotubes. Due to the thicker base layer, the nanotubes bundles become more intricate as the base layer almost covered the nanotubes' structure. Through further investigation, a spin coating rate at 2000 rpm was considered as an optimum parameter for the fabrication of $BHBT_2$:$PC_{71}BM$ composites.

During the infiltration, the solution that passes through the template nanochannels may experience two possible conditions along the process. The first possible condition is that the solution will spread and wet the nanochannels' wall, which will in turn produce the hollow nanotubes after the template dissolution. The other possible condition is the formation of nanorods, due to the existence of force interactions between the molecules (solution) during infiltration that are stronger than the adhesive force (wall). In this study, the growth mechanism of $BHBT_2$ nanostructures was rendered by the first possible condition (Figure 10 a-d). Since the optimum spin coating rate is 2000 rpm ($BHBT_2$ nanotubes), $BHBT_2$:$PC_{71}BM$ composite nanotubes were then produced at this optimum rate. Observation of hollow structure supported the prediction of wall wetting and force interaction between the wall and the solution. The wall of $BHBT_2$:$PC_{71}BM$ composite nanotubes were found to be thicker than that of the $BHBT_2$ nanotubes due to the infiltration of two different materials ($BHBT_2$ and $PC_{71}BM$).

Figures 11a and 11b show the HRTEM images of $BHBT_2$ nanotubes and $BHBT_2$:$PC_{71}BM$ composite nanotubes, respectively. $BHBT_2$ nanotubes contain only a single wall of tube whereas $BHBT_2$:$PC_{71}BM$ composite nanotubes exhibit two different regions. These two different regions corresponded to the $PC_{71}BM$ (inner) and $BHBT_2$ (outer), respectively. $PC_{71}BM$ was infiltrated into the center of the hollow $BHBT_2$, which is shown as a darker region. $PC_{71}BM$ of carbon rich materials was successfully infiltrated into the $BHBT_2$ nanotubes and wetted the inner wall of nanotubes, which led to the formation of p-n junction. If the assumption is that all of the $PC_{71}BM$ had totally infiltrated into the $BHBT_2$ nanotubes, a more effective charge transfer will take place in composite nanotubes rather than in the bulk-heterojunction. However, based on the morphological images, it was notice that not many $PC_{71}BM$ had been infiltrated into the $BHBT_2$ nanotubes. In addition to that, the nanotubes were not homogeneously constructed in terms of size. The occurrence of this phenomenon may have been due to the spin coating rate. In the spin coating process, some of the substances may have been swept away, out of the alumina template surface. Ideally, $PC_{71}BM$ substances would have to fully infiltrate the $BHBT_2$ nanotubes and to have a well-mixed of donor-acceptor interfaces in order to form the homogeneous composite nanotubes. Therefore, the composite nanotubes would provide the more effective way for the charges to be transferred into the acceptor material.

**Conclusions**

We have successfully synthesized and characterized the properties of 1,4-bis (2,2'-bithiopen-5-yl)2,5-dihexyloxybenzene ($BHBT_2$) as single material and composites. The comparison study between $BHBT_2$ thin films, $BHBT_2$ nanotubes, $BHBT_2$:$PC_{71}BM$ bulk-heterojunction and $BHBT_2$:$PC_{71}BM$ composite nanotubes emphasized on their optical, structural and morphological properties. Nanostructuring the $BHBT_2$ via template-assisted method had been successfully applied to produce nanotubes and composite nanotubes. Better quenching and improvement of charge carriers' transport was observed in $BHBT_2$:$PC_{71}BM$ bulk-heterojunction due to the poor interfaces morphology of unevenly constructed $BHBT_2$:$PC_{71}BM$ composite nanotubes. However, the enhancement of light absorption was observed in $BHBT_2$ nanotubes and $BHBT_2$:$PC_{71}BM$ composite nanotubes. $PC_{71}BM$ had been successfully infiltrated into the $BHBT_2$ nanotubes due to the wetting properties possessed by both materials, although the uniformity of infiltration is poor.

**Abbreviations**

$BHBT_2$, 1,4-bis (2,2'-bithiopen-5-yl)2,5-dihexyloxybenzene; FESEM, field emission scanning electron microscopy; HOMO, highest occupied molecular orbital; HRTEM, high resolution transmission electron microscopy; LUMO, lowest unoccupied molecular orbital; NaOH, sodium hydroxide; PA, polyacetylene; PANI, polyaniline; $PC_{71}BM$, [6,6]-phenyl $C_{71}$-butyric acid methyl ester; PCz, polycarbazole; PEDOT, poly(3,4-ethylene dioxythiophene); PF, polyfuran; PL, photoluminescence: Pth, polythiophene; PPV, polyparaphenylene vinylene; PPY, polypyrrole; XRD, X-ray Diffraction Spectroscopy

**Competing interests**

The authors declare that they have no competing interests.

**Authors' contributions**

MD carried out the experiments, performed the analysis, and drafted the manuscript. AS participated in the design of the study, performed the analysis, and helped draft the manuscript. TCH carried out the experiments, and performed the analysis. KS and RD participated in the design of the study and in the sequence alignment.

**Authors' information**

MD is a senior lecturer at the Universitas Trisakti. TCH is currently on postdoctoral fellowship at the University of Malaya. AS is the senior lecturer and KS is an associate professor at the Department of Physics, University of Malaya, while RD is a professor at the National University of Malaysia.

## Acknowledgements

The authors would like to acknowledge the University of Malaya for the project funding under the University of Malaya High Impact Research Grant UM-MoE (UM.S/625/3/HIR/MoE/SC/26), University Malaya Research Grant (RG283-14AFR), and the Malaysian Ministry of Education for the project funding under the Fundamental Research Grant Scheme (FP002-2013A).

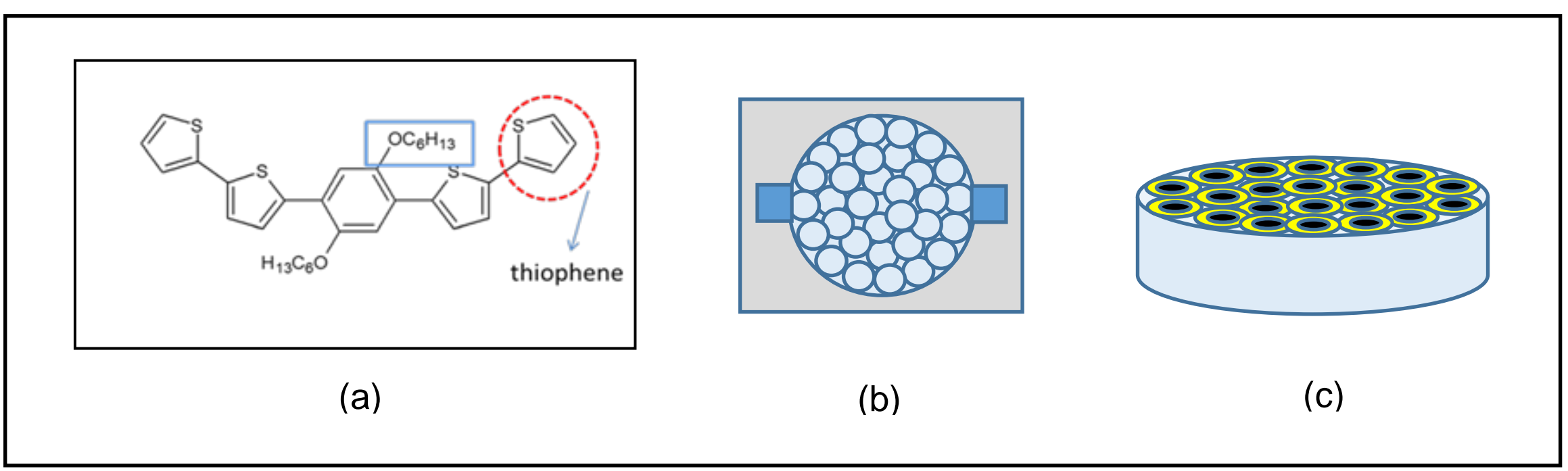

**Figure 1** (**a**) Molecular structure of $BHBT_2$. (**b**) Illustration of porous alumina template being stuck on glass substrate. (**c**) Illustration of infiltrated $BHBT_2$.

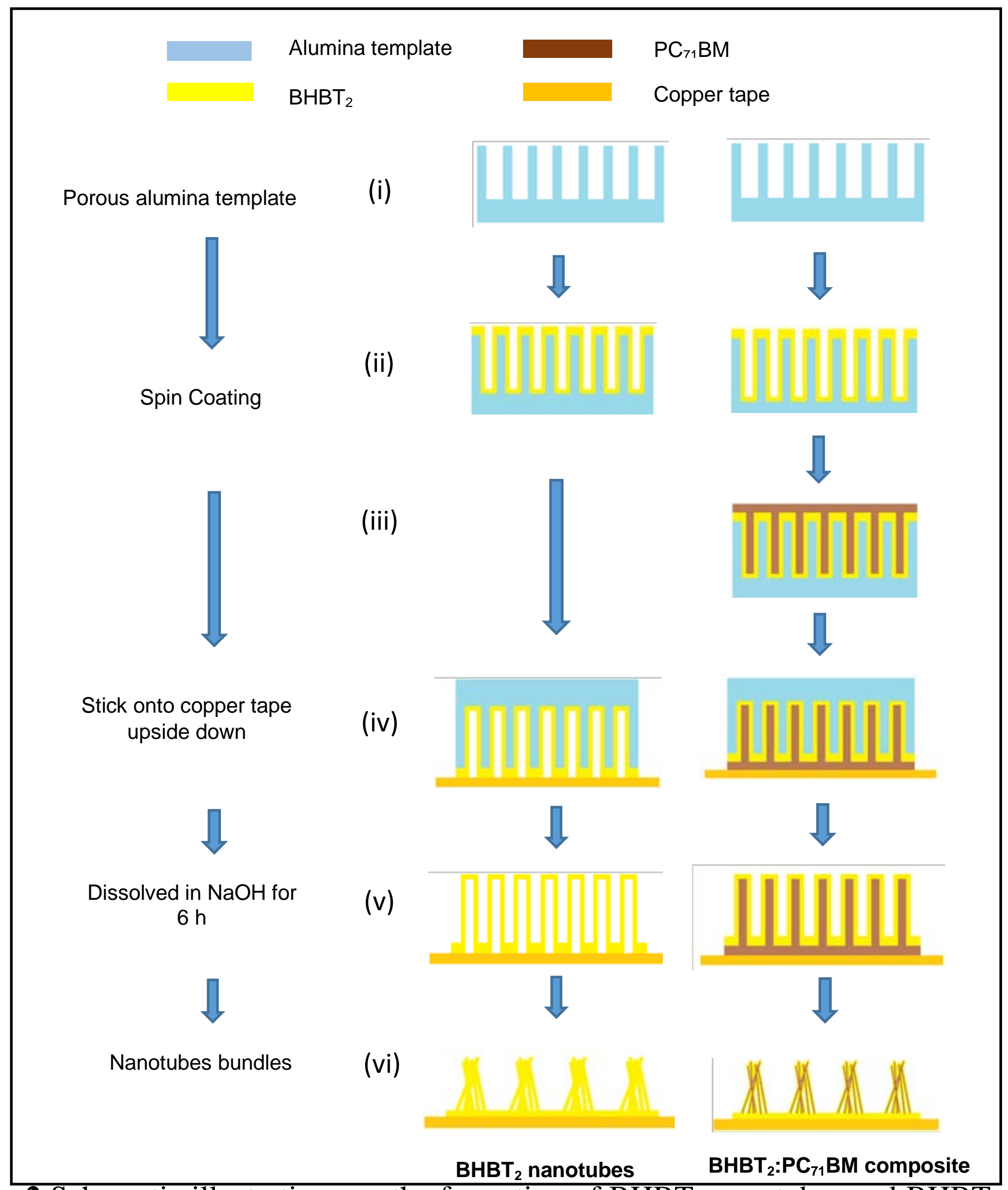

**Figure 2** Schematic illustrations on the formation of $BHBT_2$ nanotubes and $BHBT_2$:$PC_{71}BM$ composite nanotubes.

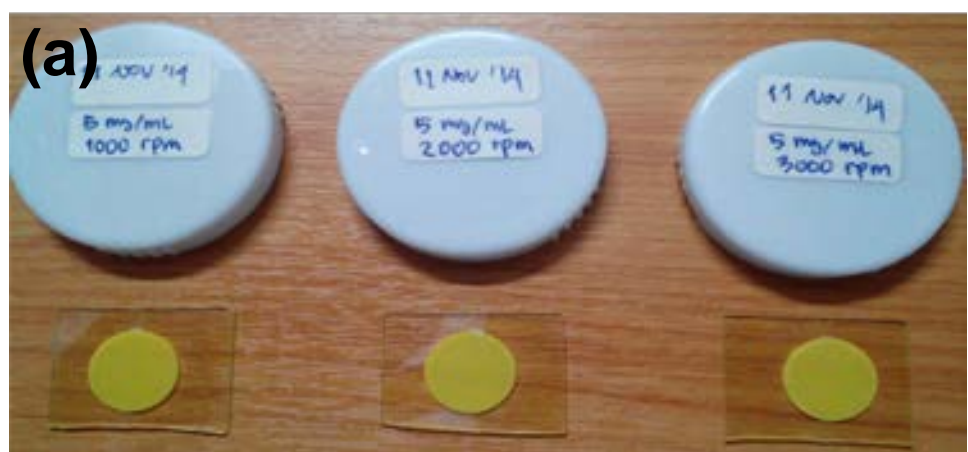

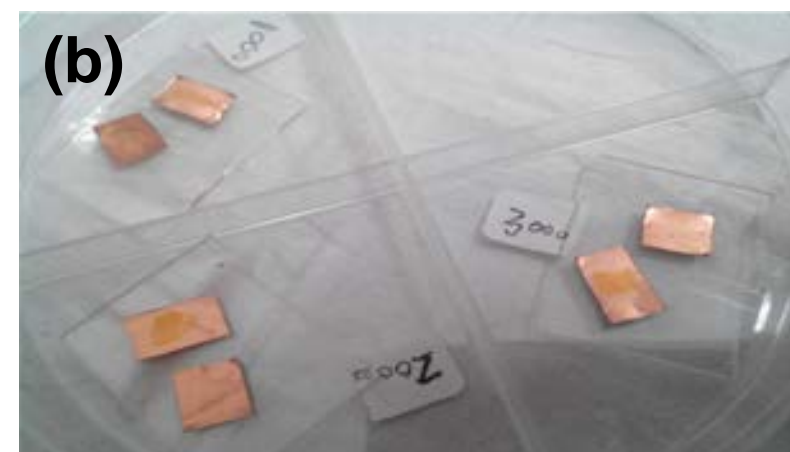

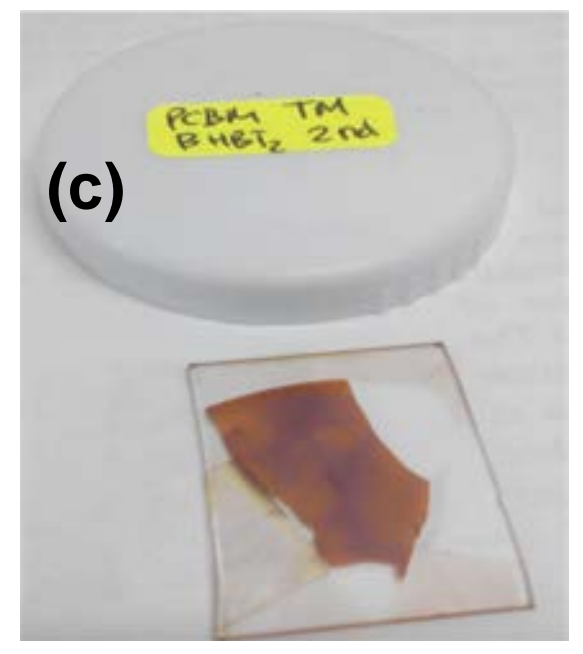

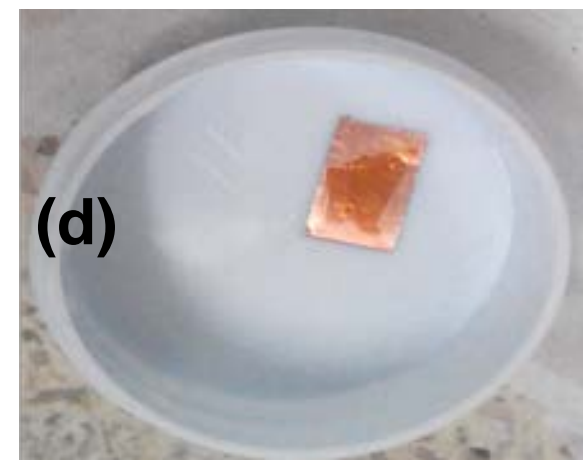

**Figure 3** (**a**) Infiltrated $BHBT_2$ that spin coated at three different rates of 1000, 2000 and 3000 rpm. (**b**) $BHBT_2$ nanotubes stick upside down on copper tape. (**c**) $BHBT_2$:$PC_{71}BM$ composite nanotubes before dissolution. (**d**) $BHBT_2$:$PC_{71}BM$ composite nanotubes stick upside down on copper tape.

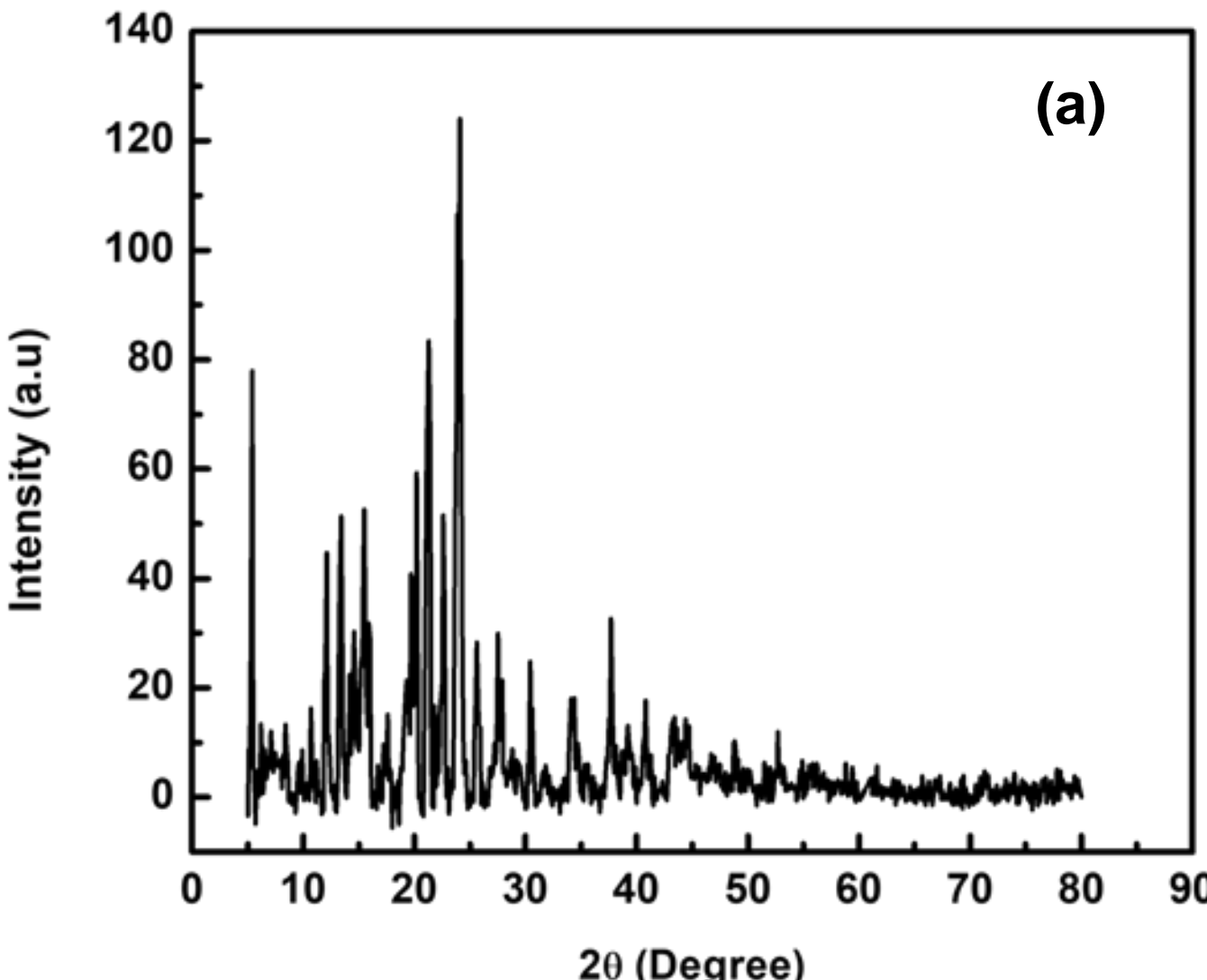

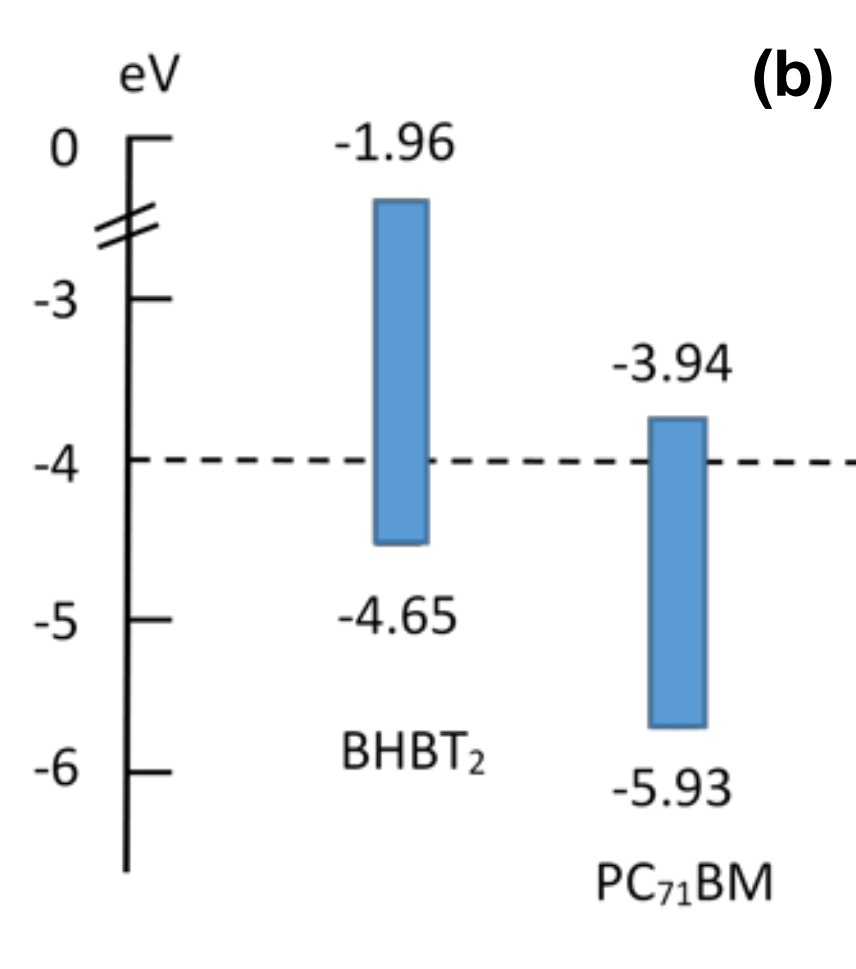

**Figure 4** (**a**) XRD measurement of pristine $BHBT_2$. (**b**) Vacuum level (energy diagram) of $BHBT_2$ and $PC_{71}BM$.

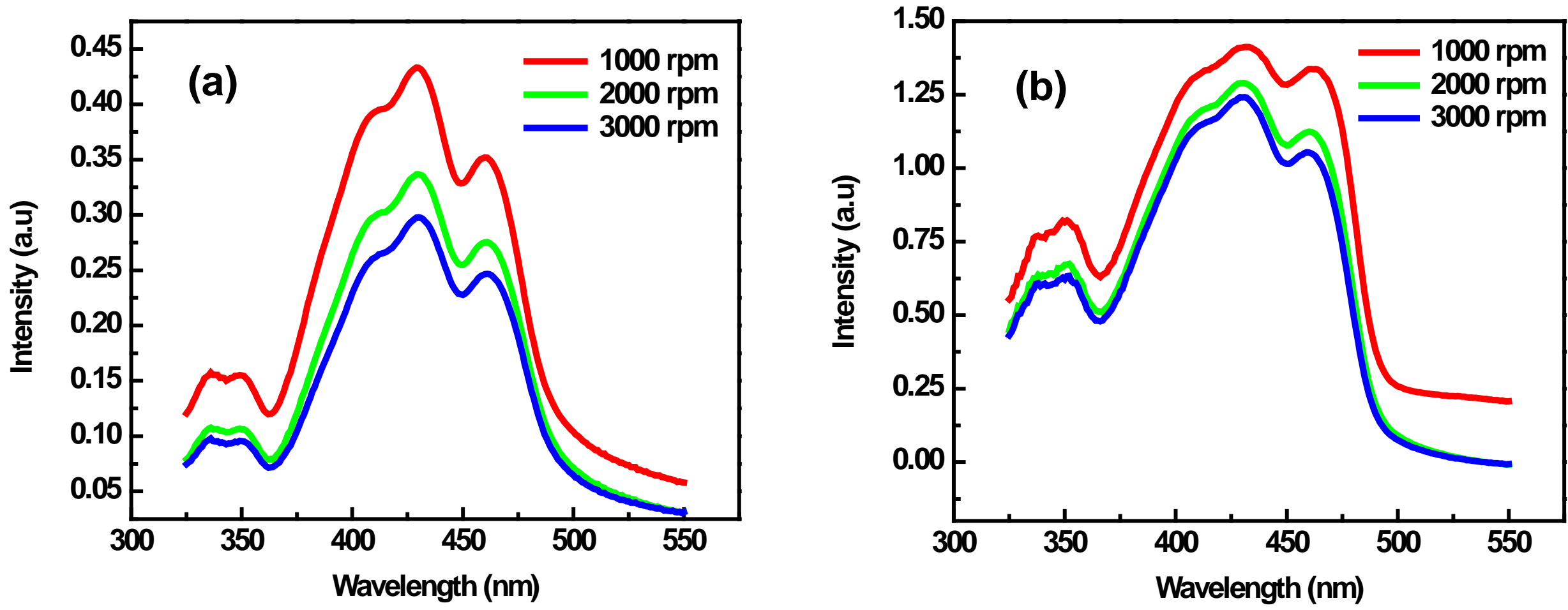

**Figure 5** (**a**) UV-vis absorption spectra of $BHBT_2$ thin films. (**b**) UV-vis absorption spectra of $BHBT_2$ nanotubes.

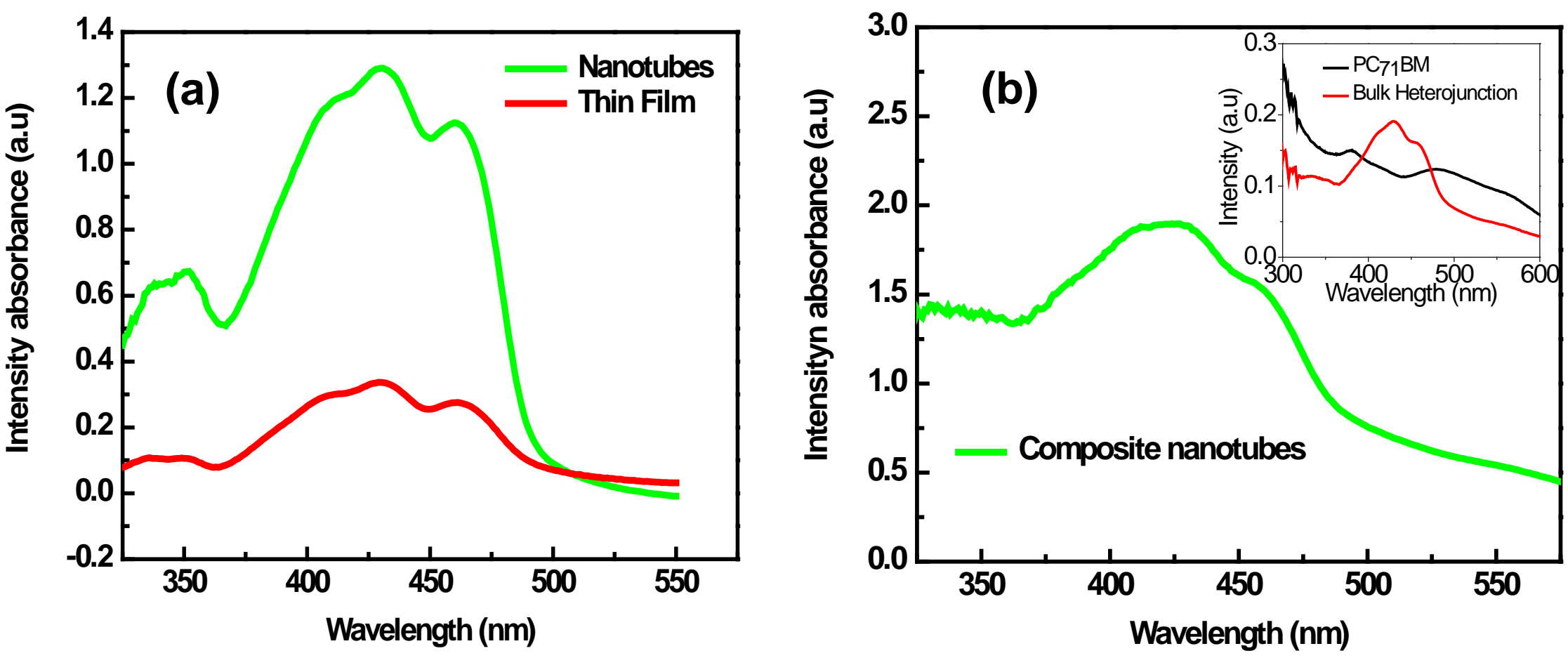

**Figure 6** UV-vis absorption spectra of (**a**) $BHBT_2$ thin films and nanotubes (**b**) $BHBT_2$ composite nanotubes and $BHBT_2$ bulk heterojunction, $PC_{71}BM$ thin films (inset).

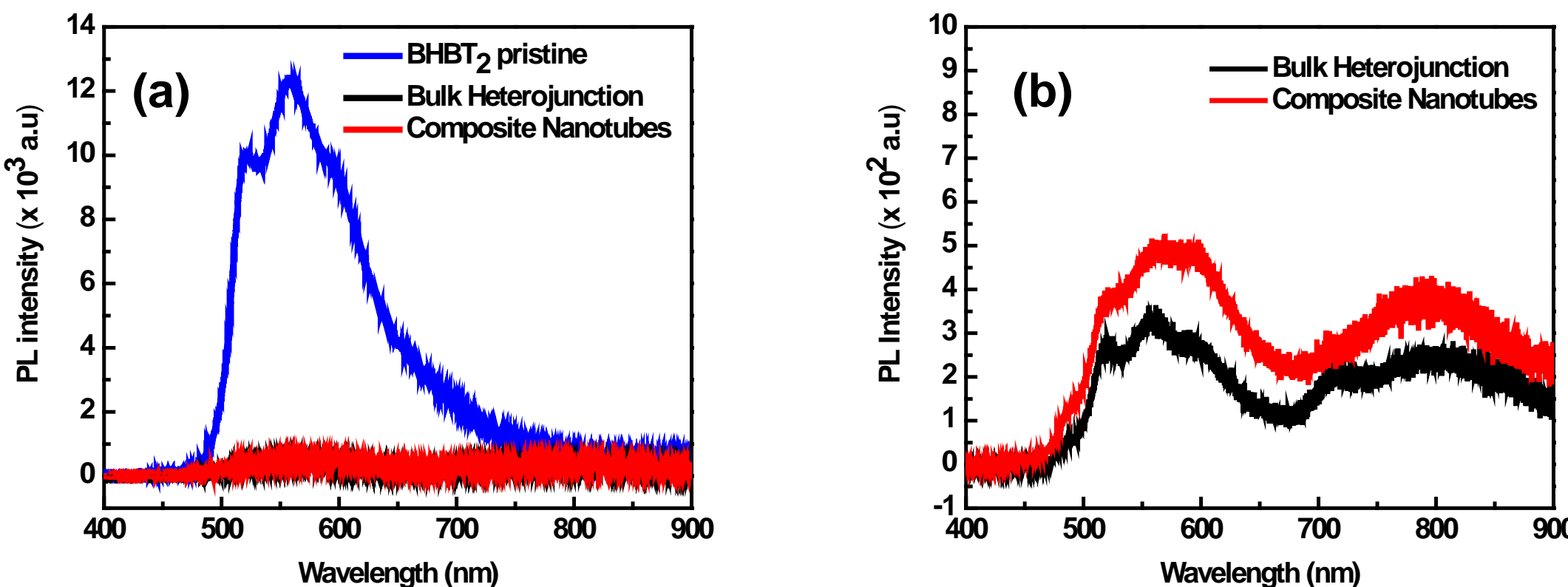

**Figure 7** (**a**) Photoluminescence spectra of $BHBT_2$ thin films, bulk heterojunction and composite nanotubes. (**b**) Photoluminescence comparison spectra between $BHBT_2$ bulk heterojunction and composite nanotubes.

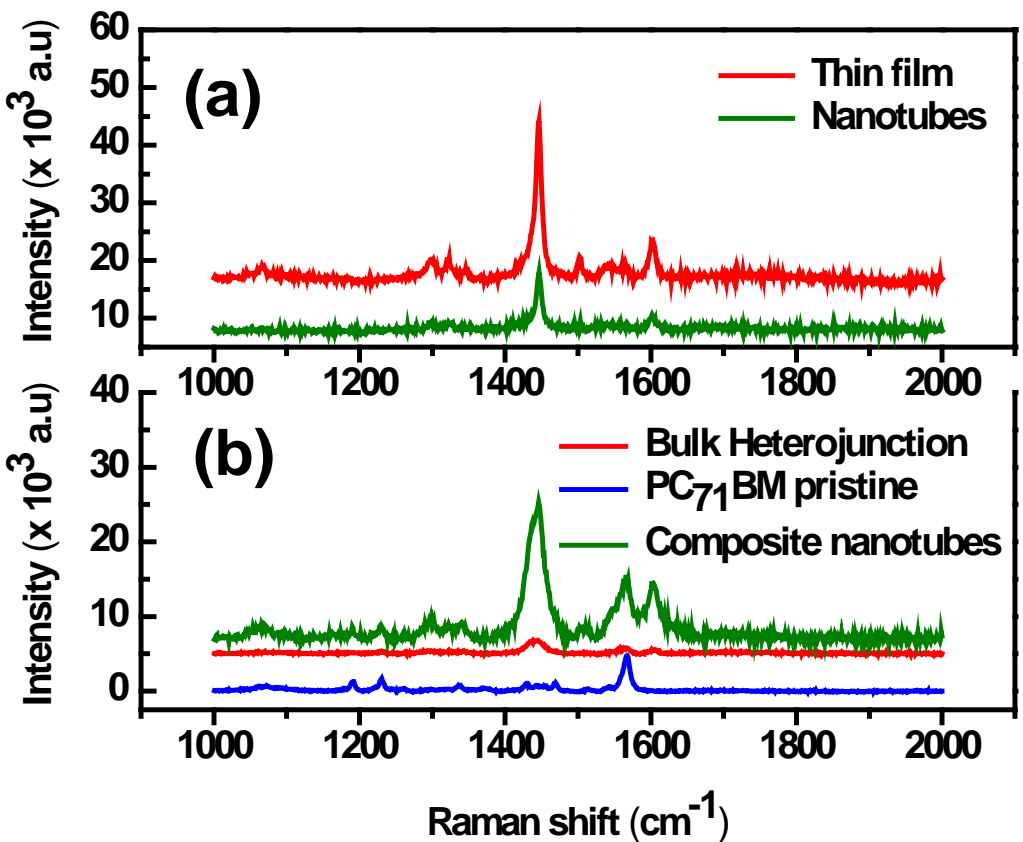

**Figure 8** (**a**) Raman spectra of $BHBT_2$ thin films and composite nanotubes. (**b**) Raman spectra of $BHBT_2$ bulk heterojunction, composite nanotubes and $PC_{71}BM$.

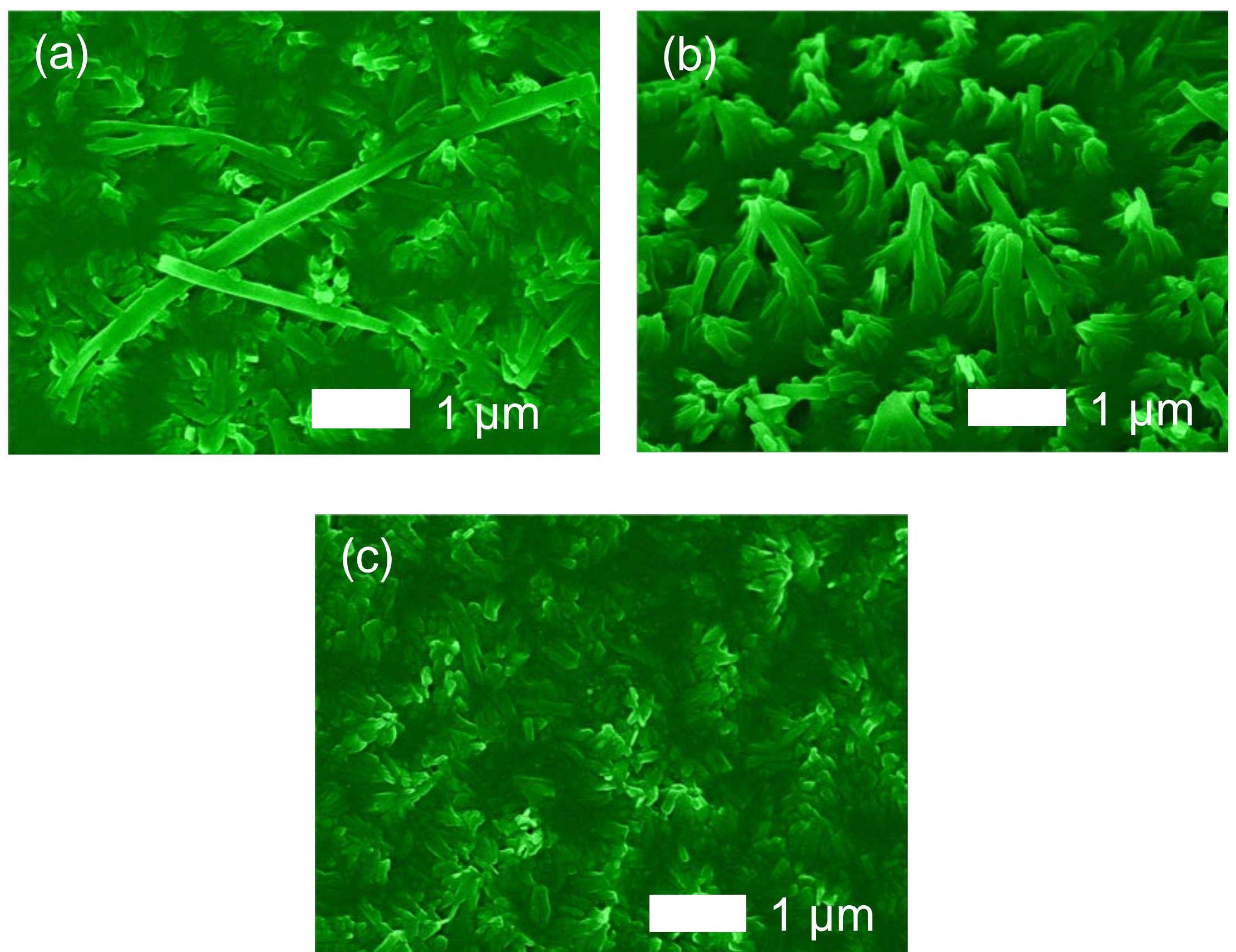

**Figure 9** FESEM images of $BHBT_2$ nanotubes at spin coating rate of (**a**) 1000 rpm (**b**) 2000 rpm (**c**) 3000 rpm.

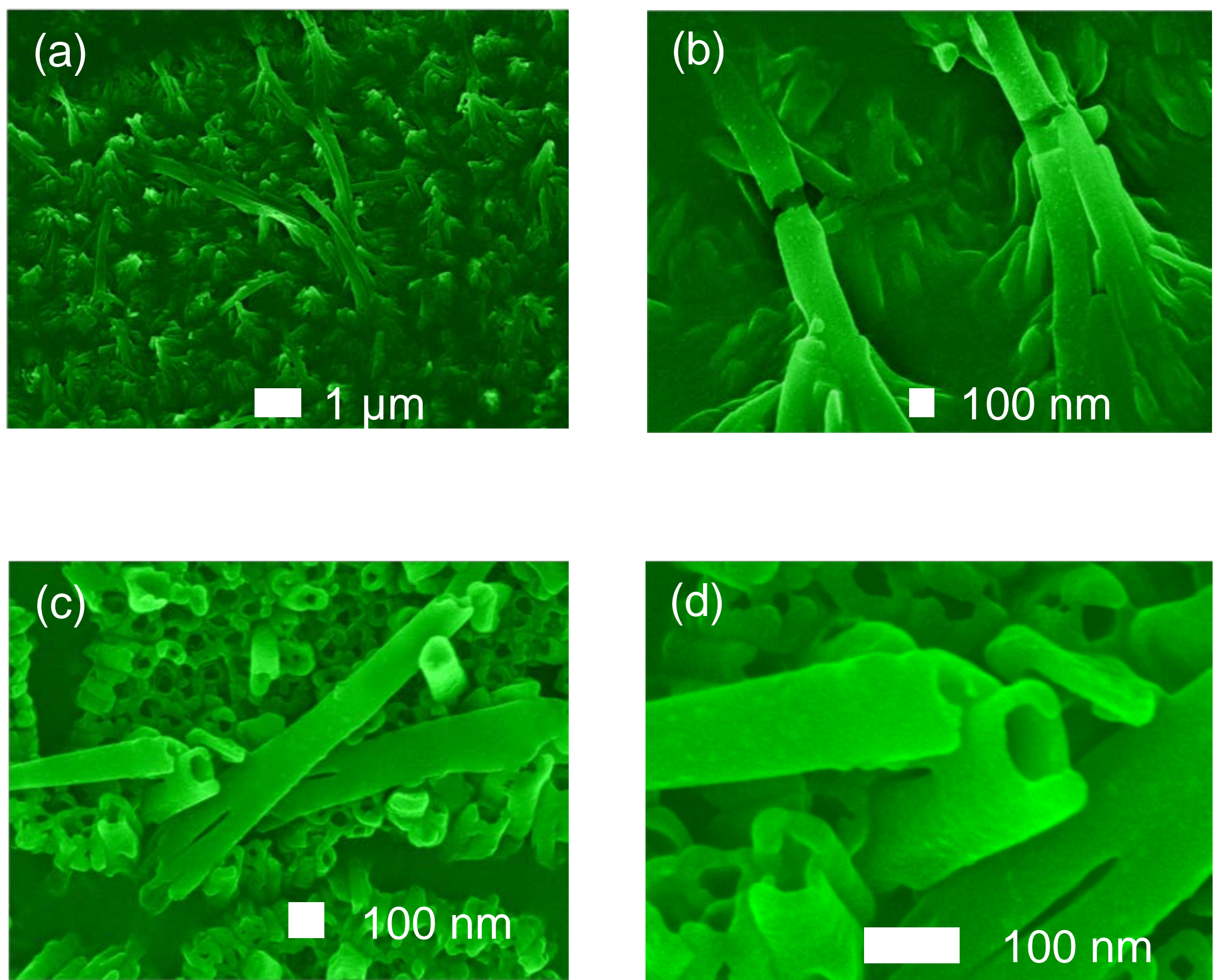

**Figure 10** (**a**) and (**b**) FESEM images of $BHBT_2$ nanotubes spin-coated at 2000 rpm. (**c**) and (**d**) FESEM images of $BHBT_2$:$PC_{71}BM$ composite nanotubes spin-coated at 2000 rpm.

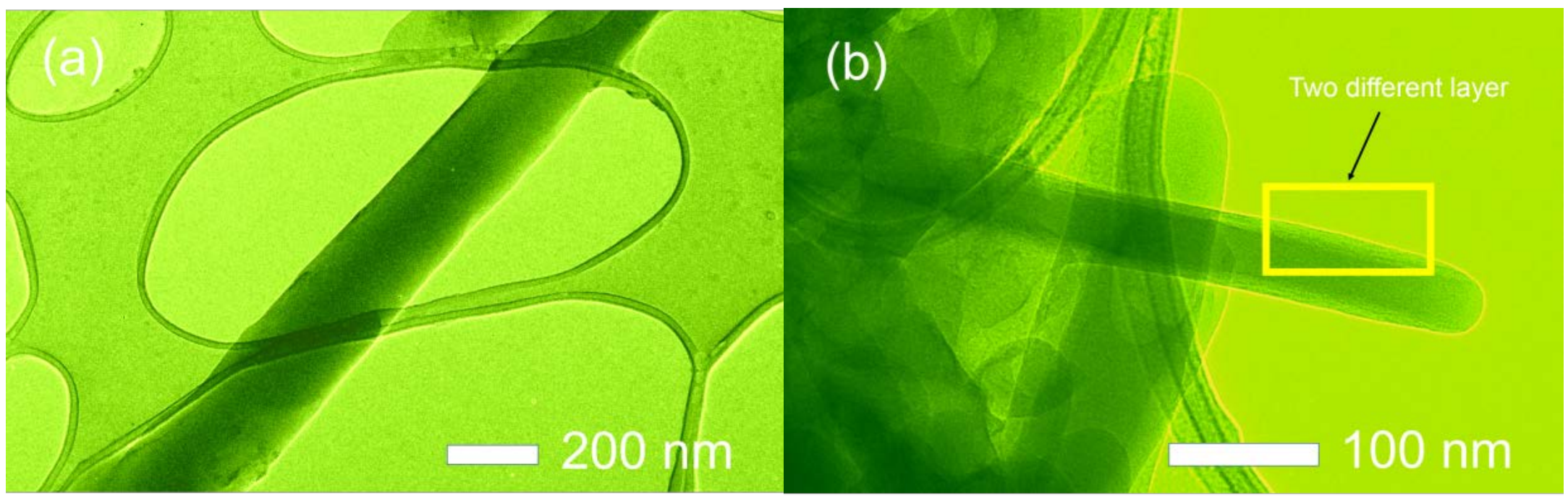

**Figure 11** HRTEM images of (**a**) $BHBT_2$ nanotubes (**b**) $BHBT_2$:$PC_{71}BM$ composite nanotubes.

**Table 1** Raman peak positions of $BHBT_2$ and $BHBT_2$ : $PC_{71}BM$ [62]

| **Raman Shift** | | | | |
|---|---|---|---|---|
| **$BHBT_2$** | | **$BHBT_2$ $PC_{71}BM$** | | |
| **Thin Film (at 2000 rpm)** | **Nanotubes (at 2000 rpm)** | **Bulk heterojunction (at 2000 rpm)** | **Composites Nanotubes (at 2000 rpm)** | **Vibrational Assignments** |
| 1067 | 1068 | 1067 | 1068 | C=S stretch Ethylene trithiocarbonate |
| - | - | 1188 | 1189 | Ring “breathing” |
| - | - | 1227 | 1231 | Ring vibration Para-disubstituted benzenes |
| 1297 | 1299 | 1295 | 1299 | CC bridge bond stretch |
| 1324 | 1325 | 1328 | 1322 | Ring vibration |
| 1344 | - | 1342 | 1338 | CH deformation |
| 1445 | 1446 | 1443 | 1443 | Ring stretch 2-Substituted thiophenes |
| 1503 | 1503 | 1507 | 1505 | Symmetric C=C stretch |
| 1544 | 1544 | - | - | C=C stretch |
| 1566 | 1564 | 1565 | 1567 | C=C stretch |
| 1601 | 1602 | 1603 | 1604 | C=C stretch |